\documentclass[journal,dvipsnames,svgnames]{vgtc}                     

\onlineid{1833}

\vgtccategory{Research}

\vgtcpapertype{Application/design study}

\title{\systemname: A Visual Analytics Tool to Explore Human Behaviour based on fNIRS in AR guidance systems}

\author{%
\authororcid{Sonia Castelo}{0000-0001-6881-3006}, 
\authororcid{Joao Rulff}{0000-0003-3341-7059},
\authororcid{Parikshit Solunke}{0009-0003-5546-0135}, 
\authororcid{Erin McGowan}{0000-0002-7565-3052}, 
\authororcid{Guande Wu}{0000-0002-9244-173X}, 
\authororcid{Iran Roman}{0000-0003-3781-7244}, \\
\authororcid{Roque Lopez}{0000-0003-3484-1783}, 
\authororcid{Bea Steers}{0009-0007-2831-6460}, 
\authororcid{Qi Sun}{0000-0002-3094-5844}, 
\authororcid{Juan Bello}{0000-0001-8561-5204}, 
\authororcid{Bradley Feest}{0009-0008-7446-5368}, 
\authororcid{Michael Middleton}{0000-0002-7964-1139}, \\
\authororcid{Ryan Mckendrick}{}, and
\authororcid{Claudio Silva}{0000-0003-2452-2295} 
}
\authorfooter{Authors are with the New York University (NYU) and Northrop Grumman Corporation (NGC). E-mails: \{s.castelo, jlrulff, pss442, erin.mcgowan, guandewu, irr2020, rlopez,  bs3639, qs2053, jpbello, csilva\}@nyu.edu. \{bradley.feest, michael.middleton, ryan.mckendrick\}@ngc.com}

\abstract{%
The concept of an intelligent augmented reality (AR) assistant has \revision{significant, wide-ranging} application\revision{s,} with potential uses in medicine, military, and mechanics \revision{domains}. Such an assistant must be able to perceive the environment and actions, reason about the \revision{environment} state in relation to a given task, and seamlessly interact with the \revision{task} performer. These interactions typically involve an AR headset equipped with sensors which capture video, audio, and haptic feedback. Previous works have sought to facilitate the development of \revision{intelligent AR} assistant\revision{s} by visualizing these sensor data streams \revision{in conjunction with the assistant's perception and reasoning} model \revision{outputs}. However, existing visual analytics systems do not focus on user modeling or include biometric data, and are only capable of visualizing a single task session for a single performer at a time. \revision{Moreover}, they \revision{typically} assume a \revision{task involves} linear progression from one step to the next. We propose a visual analytics system that allows users to compare performance during multiple task sessions\revision{,} focusing on non-linear tasks where different \revision{step} sequences can lead to \revision{success}. In particular, we design visualizations for understanding user behavior through functional near-infrared spectroscopy (fNIRS) data as a proxy for perception, attention, and memory as well as corresponding motion data (acceleration, angular velocity, and gaze). We distill these insights into embedding representations that allow users to easily select groups of sessions with similar behaviors. We provide \revision{two} case studies that \revision{demonstrate} how \revision{to use these visualizations to gain insights about}  task performance using data collected during helicopter copilot training tasks. Finally, we evaluate our approach by conducting an in-depth examination of a think-aloud experiment with five domain experts.
}


\teaser{
  \centering
  \includegraphics[width=\linewidth]{figs/teaser_v6.png}
  \caption{%
  \systemname is a visual analytics system that offers a hierarchical set of visualizations designed to analyze performer behavior in augmented reality (AR) assistance tasks by enabling multi-perspective analysis of multimodal time-series data. The \revision{Overview includes the S}catter \revision{P}lot \revision{View} (A)\revision{,} \revision{which} offers three different projections of performer sessions, revealing multidimensional clusters and patterns while \revision{enabling}  filtering and selection of sessions. The \revision{W}orkload \revision{A}ggregation \revision{V}iew (B) summarizes performers' cognitive workloads, session durations, and workload-error correlations for the chosen groups. The \revision{E}vent \revision{T}imeline \revision{View} (C) aligns multiple time series (procedures, mental workload, errors, \revision{task} phases) \revision{collected during} performer sessions along a shared time axis, enabling comparison across sessions and exploration by brushing to update linked views. The \revision{S}ummary \revision{M}atrix \revision{View} (D) facilitates analysis of procedure frequency, error proportion, overall errors, and mental state distribution within and across sessions for selected workload categories. The \revision{D}etail \revision{V}iew (E) enables in-depth exploration of individual sessions with synchronized video and time series visualizations, supporting brushing for seamless navigation and analysis.
  }
  \label{fig:teaser}
}

\graphicspath{{figs/}{figures/}{pictures/}{images/}{./}} 

\usepackage{tabu}                      
\usepackage{booktabs}                  
\usepackage{lipsum}                    
\usepackage{mwe}                       

\usepackage{mathptmx}                  
\newcommand{\systemname}{\emph{HuBar}\xspace}
\usepackage{color,soul}

\newcommand\myparagraph[1]{\vspace{4pt} \noindent \textbf{#1.}}

\newcommand\revision[1]{{#1}}

\begin{document}



\maketitle

\section{Introduction}
\label{sec:intro}

The concept of an AI-assisted task guidance system, which guides \revision{a user} through \revision{a} task using wearable sensors to detect objects and actions, is quickly shifting from science fiction to an impending reality. The potential applications of task guidance systems \revision{include} physical tasks across \revision{a wide variety of domains such as} medicine, mechanics, and military endeavors. Such a system could introduce tasks to trainees starting a new role and track their performance improvement over time\revision{, both} for their own benefit and \revision{for the} retrospective analysis of their peers. It could also serve as a second pair of eyes for domain experts, increasing task efficiency, especially during repetitive \revision{or stressful} tasks.

In recent years, enormous advancements in machine perception and reasoning, along with hardware innovations, have made it possible to begin \revision{developing} robust \revision{AI-assisted} task guidance systems~\cite{puladi_augmented_2022, nijholt_towards_2022}.
This is a complex undertaking, requiring several heterogeneous sensors and machine learning models to work together to perceive the physical environment and reason about object state changes relevant to a given task. These systems typically \revision{involve an} augmented reality (AR) headset, which superimpose\revision{s} graphics onto the performer’s real-world environment and collect\revision{s} data relevant to their behavior, \revision{(e.g.} egocentric video, audio, gaze, hand interactions\revision{)}~\cite{beams_evaluation_2022, jiang_hololens-based_2020}. Moreover, these data can be augmented with external sensors that gather information about human behavior, such as sensors to perform functional near-infrared spectroscopy (fNIRS), a popular technique \revision{for} study\revision{ing} brain activity \revision{which is} widely used to quantify mental workload\revision{~\cite{ayaz_optical_2012, solovey_using_2009}.} \revision{This performer behavior and mental workload data enable task guidance systems to adapt} task instruction\revision{s} based on the performer's mental state \revision{(for clarity, we refer to subjects using the AR system to perform tasks during a session as ``performers'' and subjects using \systemname to analyze data as ``\systemname users'')}. 



The recent \revision{development} of increasingly sophisticated AR headsets (e.g., Microsoft Hololens, Meta Quest, Apple Vision Pro) \revision{provides the} hardware necessary for AI-assisted task guidance\revision{, and has also} piqued the interest of stakeholders who could benefit from such a system. This increase in popularity \revision{has} also \revision{prompted} initiatives to collect \revision{task performance} data \revision{from} subjects with  \revision{varying expertise levels}. However, turning these data into \revision{useful insights} requires intuitive systems \revision{which enable} developers and researchers \revision{to} understand human behavior at scale and under heterogeneous constraints.

Previous efforts \revision{have} proposed approaches to explore performer actions \revision{(e.g. position and gaze)} over time using custom visualizations~\cite{castelo_argus_2024, bohus_platform_2021}. These \revision{approaches, however,} lack mechanisms to understand \revision{the performer's} mental state \revision{and how it correlates to their actions}. Furthermore, \revision{these previous works do not explore }comparison \revision{between} individuals \revision{with} different \revision{levels of} expertise \revision{at the given task}. \revision{M}ore detailed performer modeling could make AR systems more adaptable \revision{and aide} in coaching \revision{or performance report applications, especially if this performer modeling is situated in the context of data describing the surrounding environment.}


\myparagraph{Challenges in modeling performer behavior} 
To effectively model performer behavior, we must determine a method of summarizing and comparing performer behavior across sessions. This necessitates a meaningful way to compare multimodal time series data (e.g. gaze origin and direction, acceleration, angular velocity, fNIRS sensor readings) of different durations. This is a nontrivial task, especially since two performers may both successfully complete the same task by performing the same steps in different orders, or even by repeating some steps. Moreover, performer behavior modeling requires a robust method for visualizing any correlation between cognitive workload \revision{(e.g. from fNIRS sensor data)} and the sensor data streams capturing the motion of the performer. 

\myparagraph{Our Approach} We propose \systemname, a visual analytics tool for summarizing and comparing task performance sessions in AR \revision{based on} performer behavior and cognitive workload \revision{using} fNIRS, gaze, and inertial measurement unit (IMU) data. The \systemname interface is \revision{composed} of a hierarchy of \revision{four} visual components that allow the \systemname user to compare recorded task guidance sessions at varying levels of detail. At the overview level, \systemname users can compare sessions based on IMU, gaze, or fNIRS data, explore aggregated metrics for performer perception, attention, and memory workload, and select sessions of interest (see Sec.~\ref{subsection:overview}). \systemname users can then use the Event Timeline View to understand correspondences between task procedures, human errors, workload effects, and task phases for selected sessions (see Sec.~\ref{subsection:timelineView}). 
\revision{The Summary Matrix View increases the level of granularity of this analysis by showcasing how human error varies with each task procedure.}
Finally, the Detail View shows the video, IMU, and gaze data for selected portions of a given session \revision{(see Sec.}~\ref{subsection:detailsView}). All views are linked and interactive. In short, our tool facilitates post-hoc analysis of task guidance in AR through visualizations that highlight similarities and differences in performer behavior between multiple task sessions, flag human errors in task performance, and display how the performer's cognitive workload level responds to events in the physical environment. 

Our design was inspired by requirements and intermittent feedback from developers of AR systems and experts that create and evaluate these systems in the context of the Defense Advanced Research Projects Agency’s (DARPA) Perceptually-enabled Task Guidance (PTG) program \cite{noauthor_perceptually-enabled_nodate}. To summarize, \textbf{our main contributions are:}
\begin{itemize}
    \item An interactive visualization tool, \systemname, containing a hierarchy of visualizations that facilitate the exploration and comparison of performer behavior at varying levels of detail, specifically highlighting the correlations between cognitive workload, IMU, gaze, and actions during task performance. This interface was designed to enable the comparison of multimodal time series data corresponding to interleaved task procedures of differing durations. 
    \item We \revision{illustrate} the value of \systemname through two case studies that demonstrate how domain experts leverage the tool as an  after-action report and in a coaching scenario \revision{using real-world data}. 
    \item We validate our design decisions through interviews with 5 domain experts with extensive experience \revision{(collectively)} in human factors\revision{,} fNIRS, biovisualization\revision{,} neuroinformatics design\revision{, and} AR. 
\end{itemize}

This paper is organized as follows: Sec.~\ref{sec:relatedwork} reviews the relevant literature on human motion analysis based on time series, measuring workload effects in AR environments, and human behavior based on fNIRS data. Sec. \ref{sec:ocarina_dataset} describes the data.
Sec.~\ref{sec:method} specifies the requirements we aim to achieve and describes \systemname in detail, including each aspect of the visualization design. Sec.~\ref{sec:evaluation} outlines two case studies in which \systemname proves useful to experts in our chosen domain, followed by an expert interview and discussion of the feedback we received \revision{on} our system. Sec.~\ref{sec:conclusion} includes \revision{a discussion of limitations of our system,} potential future works\revision{,} and concluding remarks.

 \section{Related Work}
\label{sec:relatedwork}

\subsection{Human Behavior Analysis based on Time Series}

The analysis of human behavior using time series data from various sensors, including wearable and AR devices, is well-studied. Activity recognition, a core application, leverages data from \revision{IMUs} found in smartphones, watches, and earbuds to estimate and predict body movements over time, illustrating the potential of wearable sensors in capturing dynamic human data \cite{mollyn_imuposer_2023}.

Key to analyzing human behavior is the extraction of meaningful features from sensor data. Studies have demonstrated the use of advanced techniques, such as time series shapelets, to segment behavior activities from sensor data \cite{liu_sensor-based_2015, gong_structured_2014, qin_imaging_2020}. Fulcher’s work further underscores the significance of integrating multiple data streams for a holistic view of behavioral patterns \cite{fulcher_feature-based_2018}.


\subsection{Visualization Tools for Human Behavior Analysis based on Time Series}
Various visualization tools have been introduced to analyze and interpret sensor data for human behavior analysis. Chan et al.'s Motion Browser for analyzing upper limb movements \cite{chan_motion_2020} and Xu et al.'s ensemble of techniques for multimodal data analysis \cite{xu_finding_2020} represent significant advancements. These tools facilitate understanding of muscle coordination, behavior distribution, and interdependence among behavioral variables through sophisticated visual analytics. Notably, a study by Öney et al. provided insight into best practices for visualizing time series data collected by an AR headset using gaze data \cite{oney_visual_2023}. This system utilized both qualitative and quantitative analysis methods to provide insights into human attention and behavior in AR applications.

Together these works demonstrate a well studied area of visualization and human behavioral analysis. However, one area that these visualization techniques rarely accommodate is in human behavioral analysis with physiological measures. This is especially sparse in augmented reality tools where physiological measures are often paired with AR sensor suits to monitor an individual's activity on real world tasks.



\subsection{Insights into human performance with fNIRS}
Functional Near-Infrared Spectroscopy (fNIRS) provides physiological measurements through non-invasive tracking \revision{of} brain activity by monitoring oxygenated and deoxygenated hemoglobin levels \cite{izzetoglu_functional_2005}. FNIRS are often used as a brain-computer interface (BCI) \revision{when} movement and portability are paramount to the task being measured \cite{pinti_present_2020}. Notably, this is often the case in virtual reality, augmented reality, and real\revision{-}world tasks. Human behavior understanding can be amplified through these concentration measurements by inferring cognitive workload states in conjunction with synchronous multimodal measurements of an individual's actions and tasks. 

A key application of fNIRS is assessing cognitive workload, \revision{namely} employing behavioral models to infer workload capacity from structured tasks. These models facilitate understanding across both laboratory and real-world settings, predicting cognitive states from hemoglobin concentration data \cite{ayaz_continuous_2013, ayaz_optical_2012}. 
\revision{Research, including works by McKendrick et al., validates the cross-person and cross-task applicability of these models, demonstrating their significance in translating lab findings to practical environments \cite{mckendrick_theories_2019, maitz_towards_2023}.}


\subsection{Human Behaviour based on fNIRS}

Multimodal data, synchronously collected with fNIRS\revision{-}based cognitive workload, enriches the analysis of human behaviour, guiding the design of more responsive and adaptive real\revision{-}world systems. Mark et al. provides a comparison study which incorporates various brain-body measures to offer insights into cognitive processes over time \cite{mark_mental_2024}. Similarly, Yuksel et al. demonstrate how adjusting task difficulty based on cognitive load readings and behavioral measurements can significantly improve learning efficiency, as seen in \revision{their} adaptive piano training program \cite{yuksel_learn_2016}.

In high-stakes environments like aviation, fNIRS is often used for monitoring cognitive workload and fatigue, offering insights into pilot engagement and decision-making processes \cite{ayaz_continuous_2013, ayaz_optical_2012}. Various studies \cite{verdiere_detecting_2018, dehais_monitoring_2018, yuan_recognition_2024, mark_neuroadaptive_2022} highlight \revision{the role of} fNIRS in evaluating pilot performance in varied scenarios, including real and simulated flights, thereby showcasing the modality's adaptability and effectiveness in critical applications. On top of this, many of these studies rely upon the need for multiple modes of synchronous sensor data \revision{in addition to} fNIRS physiological measurements to understand human behavior.

Integrating fNIRS with other synchronous time series modalities enriches behavioral analysis, allowing for a nuanced understanding of human cognition. This \revision{prompts} the need for visualization tools to assist behavioral specialists in interpreting complex interconnected time series datasets, \revision{specifically tools which} link physiological measurements to specific behaviors and decisions.

\subsection{Measuring workload effects in AR environments}


 
The vast majority of previous studies \revision{about} cognitive workload effects in AR environments focus on measuring the impact of using an AR headset on the wearer's mental workload during a task \cite{atici-ulusu_effects_2021, qin_eeg-based_2023, maag_measuring_2023}, rather than measuring the cognitive workload of a person who just so happens to be performing the task in an AR environment. 
\revision{Caarvida \cite{achberger_caarvida_2020} and AutoVis \cite{jansen_autovis_2023}, for instance, provide tools to explore automotive test data, but do not draw correlations between cognitive load and performer actions and errors.}

More recent AR studies involve interfaces that evolve as a function of the individual's workload state, requiring real-time and multimodal behavioral analysis \cite{maitz_towards_2023, galati_exploring_2021}. This brings a new set of challenges to visualization tools. We need tools that can generalize to many environments, run in real time, and visualize many synchronous streams of data. Galati, Schoppa, and Lu implement a visualization tool in their AR pipeline \cite{galati_exploring_2021}. This tool features an interactive exploration of \revision{user movements with respect to} raw fNIRS signals, allowing experts to compare and identify areas of cognitive activity in the raw signal. However, this tool is tailored to handle data from this specific study\revision{.} \revision{F}urthermore\revision{,} it is visualizing raw fNIRS data instead of classified workload states. While this is \revision{useful} for neuroscientists \revision{who want} a better understanding of spatial brain data, it is not as helpful for the \revision{broader} study of human behavior\revision{, particularly by AR system developers who may not have a neuroscience background}.

To the best of our knowledge, there are no such generalized tools that \revision{aide} human behavior specialists \revision{in} analyz\revision{ing} these many synchronous streams of data for augmented reality systems. Such a tool will improve both the efficiency of analysis as well as the conclusions that can be drawn from human behavior data.

\section{Ocarina dataset}
\label{sec:ocarina_dataset}

The Ocarina dataset\revision{, collected by NGC as part of the DARPA PTG program,} consists of data from simulated UH-60V helicopter copilot sessions, totaling approximately 3TB. It encompasses data from 7 participants and 33 sessions. Each participant participated in one to eleven sessions. Every session corresponded to a specific task scenario, with each task comprising multiple non-sequential procedures that could be completed in different orders. 
\revision{After recording, these procedures were extracted from the mission logs and designated alphabetical names from ``a'' to ``f''. This subset of six procedures was chosen from the nine possible ones as they accounted for $98.5\%$ of all procedure occurrences.} 
Each unique task scenario is identified by a distinct ``trial ID'' within the dataset, except for trials 2, 10, and 23, that represent the same task.  

\myparagraph{Participants}
Data collection for the Ocarina dataset involved 7 participants. Three participants had previous piloting experience. No pilot had direct experience with the specific UH-60V cockpit. Three participants were engineers with experience developing the software for the UH-60V cockpit. Each had varying levels of experience with the system, but all were familiar with the cockpit. Two of the engineers were highly\revision{-}versed in the logic of the system and had directly developed several of its capabilities. The seventh participant was a computer science professor at a large North American university. In total, participants completed 47 flights.

\myparagraph{Data Collection Protocol}
Participants were seated in front of a physical recreation of the UH-60V cockpit, with mission computers that replicate flight systems and simulate flight routes and in-flight events. They were outfitted with recording devices, including an fNIRS neuroimaging system. Additionally, a Microsoft Hololens 2 was placed atop the fNIRS device. The Hololens collected audio, video, IMU data containing accelerometer, gyroscope, and magnetometer readings, eye tracking data consisting of \revision{3-dimensional vectors for} gaze origin positions and directions, and hand tracking data consisting of 26 joint points for each hand.
Throughout the data collection process, participants performed full flights, comprising pre-flight and flight phases. To advance to the flight phase, participants needed to complete nine procedures during the pre-flight phase. The mission computer logs recorded 
all physical interactions the participants made with the simulator during each trial.

\myparagraph{fNIRS Workload Classification}
Predictions of \revision{the} participants' cognitive workloads, spanning working memory, attention, and perception, are derived from the fNIRS time series \revision{measurements of} hemoglobin concentration. \revision{These measurements, captured at a frequency of 10 Hz, include raw light intensities, HbO (oxyhemoglobin), and HbR (deoxyhemoglobin) concentrations. These measurements serve as inputs for dedicated classifiers; at each time-step, three mental states are specified: perception, attention, and workload. Each mental state is classified as either ``optimal'', ``overload'', or ``underload'' with an associated classification confidence \revision{(we elaborate upon the interpretation of these classes in the following paragraph)}. E}ach workload category has its own classifier, which \revision{runs} concurrently during recording \revision{sessions}. The classifiers are generalized mixed effects models trained \revision{on} data gathered from a previous test bed study that showed cross-task and cross-participant transfer \revision{\cite{mckendrick_theories_2019}}. The performer's classified mental state is first base-lined before recording and is not shown to the performer during collection.

Working memory capacity \revision{(which we hereafter refer to as \textit{memory})} is an individual's ability to retain and manipulate information during task execution. \textit{Attention} pertains to the individual's capacity to concentrate on specific tasks selectively. \textit{Perception} is the individual's ability to interpret stimuli, both visual and auditory. The Ocarina dataset categorizes these cognitive facets into three states: optimal, overload, and underload. An optimal state denotes a balanced cognitive load conducive to task performance. An overloaded state suggests a cognitive burden exceeding an individual's capacity, potentially impairing the incorporation of new information~\cite{young_state_2015}. Conversely, an underloaded state indicates a cognitive engagement below the individual's capacity, which may result in diminished focus~\cite{young_state_2015}. It is critical to monitor multiple synchronous information streams alongside cognitive state data, as the presence of an overloaded or underloaded state does not invariably correlate with decreased task performance.

\section{Method}
\label{sec:method}
\subsection{Domain Requirements}
\label{subsec:domain_requirements}

\revision{The design requirements of \systemname were defined during a year-long collaboration with researchers, who are coauthors of this paper, actively developing an end-to-end task guidance system to support pre-flight procedures. In addition, we conducted multiple interviews with data scientists who actively work in fNIRS data analysis and data visualization to validate our design choices.}

\begin{enumerate}[start=1,label={[\bfseries R\arabic*]}]

\item \label{req:performer_behavior} \myparagraph{Performer behavior overview} The experts stressed the importance of having the ability to visualize all performers' behavior in a single view. This helps trainers categorize performer expertise based on their behaviors across sessions. Trainers want to identify performers, for example, who may need additional training. Additionally, the trainers would like to know specific procedures where a certain performer excels \revision{or} struggles. Finally, the trainers would like to detect and investigate clusters of similar sessions or performers. We propose a combination of the \revision{S}catter \revision{P}lot and \revision{S}ummary \revision{M}atrix \revision{V}iews to tackle this requirement. 
\vspace{-.15cm}

\item \label{req:align_multiple_sessions} \myparagraph{Aligning and comparing multiple sessions} Visualizing and comparing multiple sessions, each with multiple attributes based on time series data\revision{,} can be challenging when there is no implicit sequentiality \revision{(}as is the case with the Ocarina dataset\revision{)}. A time-aligned view of the procedures, errors, workload information, and \revision{task}
phase information would help trainers discern important landmarks within a particular session. Furthermore, this would enable trainers to compare landmarks and timestamps across multiple sessions and performers. To tackle this problem, we propose the \revision{E}vent \revision{T}imeline \revision{V}iew that combines the various streams of time series data along a common time axis\revision{,} enabling seamless comparison across sessions. 
\vspace{-.15cm}

\item \label{req:correlations} \myparagraph{Compare fNIRS data across sessions and visualize correlations between fNIRS data and performer behavior} The experts stated they were interested in visualizing and comparing fNIRS data across different performers at different levels of granularity. They would like to investigate fNIRS summaries for subjects across all sessions, while also being able to drill down \revision{in}to a single session and make comparisons between sessions. Furthermore, experts were particularly interested in understanding correlations between the mental states (\revision{o}verload, \revision{u}nderload, and \revision{o}ptimal) of performers for each workload category (\revision{a}ttention, \revision{p}erception, \revision{m}emory) \revision{and} errors made during sessions. Finally, experts would like to know the correlations between mental states and specific procedures. To this end, we propose the Workload Aggregations in the Overview, along with detailed fNIRS data for individual sessions in the \revision{E}vent \revision{T}imeline and \revision{S}ummary \revision{M}atrix \revision{V}iews. Furthermore, we display the correlations between errors and the mental states for \revision{both} sessions and performers. 
\vspace{-.15cm}

\item \label{req:detailed_view} \myparagraph{Detailed visualization of performer behavior} To uncover associations between performer behavior, fNIRS predictions, actions, and errors, trainers need to explore individual sessions in great detail. Trainers would greatly benefit from being able to analyze data from the IMU and gaze sensors \revision{in conjunction} with the \revision{egocentric} video \revision{captured by the AR headset}, \revision{as it would} allow them to detect patterns and establish connections between the various data streams. To meet this requirement, we propose \revision{the D}etail \revision{V}iew\revision{, which includes} interactive visualizations for IMU and gaze data linked to the session video. 

\vspace{-.15cm}
\end{enumerate}

\subsection{Visualization Design}

 We employed the ``overview-first, zoom and filter, details-on-demand'' strategy \cite{shneiderman_eyes_2003} as our guiding principle while formulating the visual design, with the goal of ensuring a user-centric approach that facilitates efficient exploration and comprehension of the multimodal data associated with performer sessions. \revision{The resulting tool,} \revision{\systemname, is composed} of four linked interactive views: \revision{the O}verview (Figure \revision{\ref{fig:teaser}(A) and \ref{fig:teaser}(B)}), the \revision{E}vent \revision{T}imeline \revision{View}, \revision{the S}ummary \revision{M}atrix \revision{View}, and \revision{the D}etail \revision{V}iew \revision{(Figures~\ref{fig:teaser}(C), \ref{fig:teaser}(D), and \ref{fig:teaser}(E), respectively)}.

\subsubsection*{Overview}
\label{subsection:overview}

The \revision{O}verview consists of two sub-views: \revision{the S}catter \revision{P}lot \revision{V}iew \revision{shown in Figure~\ref{fig:teaser}(A), which allows users to select sessions based on various features,} and \revision{the W}orkload \revision{A}ggregation \revision{V}iew \revision{shown in Figure~\ref{fig:teaser}(B)}, which \revision{displays} \revision{cognitive} workload and session duration information based on user selection. 

\myparagraph{Scatter Plot} \revision{The \revision{S}catter \revision{P}lot \revision{View} shown in Figure~\ref{fig:teaser}(A) serves as the starting point for exploration in \systemname. The \revision{Scatter Plot View} categorizes sessions to \revision{facilitate} the identification and comparison of similar sessions or outliers \ref{req:performer_behavior}. The user can adjust the scatter plot to represent only the performers' physical activity (IMU, gaze) or brain activity (fNIRS) by selecting the desired data stream. Below, we detail the process of transforming time series \revision{data} into 2D scatter plot points. Each point in the plot represents a session, and different symbols represent either trials or subjects, depending \revision{upon user} selection. The user can lasso-select these symbols, and \revision{remaining} views will update accordingly.} 
\revision{Users can also opt to display only the types of tasks which appear most frequently in the dataset (e.g. ``top 10'').}

\myparagraph{Brain and Physical Activity} \revision{The user can toggle whether the points in the Scatter Plot View represent IMU, gaze, or fNIRS data. Toggling to IMU or gaze enables the user to select sessions based on the performer's physical activity throughout the session; IMU data represents the performer's body movement, whereas gaze data represents the displacement of their visual attention. To compare these time series, we transform them into 2D vector representations. We chose to do this using a shapelet-based \cite{ye_time_2009} technique due to its ease of use and robust implementation through the \texttt{tslearn} library \cite{tavenard_tslearn_2020}. Although this algorithm requires some preprocessing of the data, such as normalizing time series to the same length, our system is agnostic of the technique. Other approaches (e.g. TS2Vec \cite{yue_ts2vec_2022}) could be substituted in cases where the normalization of the time series could hide important information about the sessions.}

\revision{In contrast, the user may toggle the Scatter Plot View to show fNIRS time-series data,  shifting the focus of the exploration to brain activity throughout the sessions. A similar process could be applied to generate the projection of points based on fNIRS data as was used for the IMU and gaze data. However, in the current implementation of \systemname, we transform the raw fNIRS signal using the workload classification models described in Section \ref{sec:ocarina_dataset} before generating the 2D vectors that are ultimately rendered in the Scatter Plot View.}

\myparagraph{Workload Aggregation View}
\revision{We showcase the proportion of time spent in each mental state (overload, optimal, and underload) across the three workload categories (memory, attention, and perception) aggregated by the selection made in the scatter plot view (Figure \ref{fig:teaser}(B)). To convey the ordinality of these mental states, we employ a sequential red color scale where light red represents underload, a medium shade indicates optimal conditions, and dark red signifies overload} \includegraphics[width=0.6cm,height=0.7em]{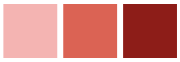}.

Furthermore, we present the error contribution linked to each mental state across all categories for each group. This metric is crucial\revision{,} as it presents the correlation between performers' errors and their respective mental states for the three categories. \revision{To display this information effectively, we opted for an aligned position against a common scale plot. This choice facilitates easy comparison and identification of data of interest while saving vertical space and reducing visual clutter in the overall system. We intentionally avoided using a bar chart-based plot to minimize potential confusion with the workload bar chart. To avoid overlapping from identical correlation scores, we adapt the scale accordingly.}
Given a \revision{selected} group of sessions $g_i$, we estimate the error correlation using Pearson Correlation (PC) between $e$, the error duration, and $s$, the state duration of each workload for the three categories (optimal, overload and underload). The sample points for these variables were collected for each procedure measured in seconds. Last, we highlight the average session duration for the selected groups \ref{req:correlations}. 

\subsubsection*{Event Timeline View}
\label{subsection:timelineView}
In the \revision{E}vent \revision{T}imeline \revision{V}iew \revision{shown in Figure~\ref{fig:teaser}(C)}, we coalesce data from four different data streams recorded during performer sessions into a unified, time-aligned visualization for each selected session. These sessions are organized by trial ID or subject ID, \revision{as chosen in the \revision{S}catter \revision{P}lot \revision{V}iew.} Duration is represented along the x-axis, \revision{beginning} at zero for each session.

\revision{Task steps or procedures are visualized using horizontal bars that extend for the duration of each session. Segments within these bars are color-coded according to the ongoing procedure at the corresponding timestamp. We excluded shades of red from this color scale to prevent any conflict with the scale used for the workload variable.} Furthermore, we have an error bar that employs black segments to indicate errors at their corresponding timestamps. Next, we have the workload bar, \revision{where segments illustrate the performer's mental states (underload, optimal, overload) for the chosen workload category.} Furthermore, the model confidence score for its predicted mental state is depicted using a line within the bar graph. \revision{Finally, we have the task phase indicator, which may be used to group task steps or procedures (e.g. in the case of the Ocarina dataset, this is where we use ``PF'' and ``FL'' to denote the pre-flight and flight stages, respectively).} 

The rationale for aligning the various data streams along the time axis is multi-faceted. First, employing a unified time scale across all selected sessions facilitates convenient evaluation of their respective durations. Moreover, it allows \systemname users to compare the performers' mental states and the errors committed across different sessions. In addition to inter-session evaluation, the design facilitates intra-session evaluation by enabling \systemname users to promptly identify error occurrences and establish potential correlations between errors and the corresponding procedures, mental states, and flight phase \ref{req:align_multiple_sessions} \ref{req:correlations}.

\revision{Consider the scenario where the \systemname user wants to investigate a particular session. To  \revision{do this}, the\revision{y simply brush the Event} \revision{T}imeline \revision{V}iew along the time axis. \revision{This updates the S}ummary \revision{M}atrix \revision{V}iew\revision{, which} employs transparency and opacity to highlight the procedures involved in the brushed section. \revision{This also updates the D}etail \revision{V}iew to display \revision{egocentric} video and sensor data corresponding to the brushed timestamps, enabling \systemname users to see the pilot's perspective and sensor readings for the selected period.}

\subsubsection*{Summary Matrix View}
\label{subsection:summaryMatrixView}

\revision{The interviewed experts showed great interest in comparing errors, mental states, and prevalence of procedures within a session as well as across sessions \ref{req:performer_behavior}. However, due to the non-linear nature of \revision{the procedures performed in many tasks}, it \revision{can be} challenging to discern these nuances when \revision{the data is} visualized sequentially.} To address these challenges, we propose the \revision{S}ummary \revision{M}atrix \revision{V}iew (Figure \ref{fig:teaser}\revision{(D))}, which complements the \revision{E}vent \revision{T}imeline \revision{V}iew to give a \revision{more nuanced} picture of performer data. \revision{It includes} pie charts for every procedure, where \revision{chart} radius corresponds to \revision{procedure} prevalence. The pie charts are shaded black and gray to based on the proportion of errors \revision{(represented by the black slice)} within the corresponding procedure. \revision{Pie charts are employed here specifically to communicate two proportions simultaneously: (1)} the proportion of a particular procedure in the duration of a session \revision{and (2)} the proportion of error within each procedure for the session. This allows the \systemname user to compare procedures and associated errors with different procedures for the same session (horizontally), as well as with the same procedure for different sessions (vertically) \ref{req:performer_behavior}. 

\revision{In addition to the pie charts, we show the proportion of errors and the distribution of mental states for the chosen workload category for each session. The provided checkbox can be used to show or hide the error contribution for mental states within the selected workload category \ref{req:correlations}.} Since this error correlation corresponds to the individual session, we used the regular PC to calculate it (similar to the Workload Aggregation \revision{V}iew). \systemname users can select the desired category either through the dropdown in this view, or by clicking the corresponding category label in the \revision{W}orkload \revision{A}ggregation \revision{V}iew. Additionally, transparency is used to fade out the non-selected categories in the workload aggregations view. \revision{This design decision is intended to help \systemname users retain focus by reducing visual clutter.} 
\revision{Furthermore, the pie charts provide a\revision{n on-hover} tooltip which} display\revision{s} the correlation between errors $e$ and the mental states $s$ within the corresponding procedure $p$ ($p$ is a vector where $p_i=1$ if the procedure $i$ is the procedure in question and $p_i=0$ otherwise). We calculate these values using Partial Correlation \cite{baba_partial_2004}. 
Let $r_{se}$ be the correlation between $s$ and $e$;  $r_{sp}$, the correlation between $s$ and $p$; and $r_{ep}$,  the correlation between $e$ and $p$. The Partial Correlation  is computed as:  $r_{se,p} = \frac{r_{se} - r_{sp} \cdot r_{ep}}{\sqrt{\left ( 1-r_{sp}^{2} \right )\cdot \left ( 1-r_{ep}^{2} \right )}}$. 

\subsubsection*{Detail View}
\label{subsection:detailsView}
\revision{One of the major requirements expressed by the interviewed experts was the ability to investigate individual sessions and observe performer (e.g. pilot) actions in detail ~\ref{req:detailed_view}. The \revision{D}etail \revision{V}iew was designed to meet this requirement (Figure~\ref{fig:teaser}(E)). The video view plays the \revision{egocentric} video from the performer's perspective corresponding to the brushed timestamps. \revision{Below} this, we \revision{use line plots to} visualize data from the IMU and gaze sensors, capturing the performer's \revision{body and eye motion} over time.} The \systemname user can switch between the variables corresponding to the IMU and gaze sensors using \revision{their} respective dropdowns. Finally, the segmented bar graph depicts the mental states for the chosen workload category throughout the session. The time window brushed in the \revision{E}vent \revision{T}imeline \revision{V}iew is highlighted in all three visualizations within this \revision{D}etail \revision{V}iew. All three visualizations can be brushed, similar to the \revision{E}vent \revision{T}imeline \revision{V}iew. Moreover, the brushes are all synchronized with each other and with the video player, facilitating seamless navigation and exploration of sessions. Aligning the IMU and gaze data with the video, workload information, and procedures enables the \systemname user to identify procedures with high levels of human motion, establish associations between motion levels and mental workload as well as errors, and navigate to these regions of interest in the video by simple brushing \ref{req:detailed_view}.

\section{Evaluation}
In the aviation industry, pilots often experience mental states of overload or underload which can have immediate consequences such as heightened stress, monotony, mental exhaustion, or fatigue. In addition to posing significant risks to flight safety, these short-term effects, if not addressed, can escalate into long-term issues such as psychosomatic or mental health disorders. In the following case studies, we describe how a pilot trainer and an AR guidance system developer can use \systemname to recognize and evaluate the factors contributing to overload or underload mental states in copilots.

\label{sec:evaluation}
\subsection{Case Study 1: Unraveling the Triggers of Mental Underload in Co\revision{p}ilots }

To showcase how the \systemname system supports  effective exploration of \revision{task} sessions within real-world contexts, we present a case where a pilot trainer utilized the system. The trainer aimed to identify instances during flight procedures where a copilot might experience an underloaded mental state, discern potential causes behind such occurrences, and extract valuable insights from the data. The underload mental state is particularly concerning during a flight as it may indicate that the \revision{copilot} is overly relaxed or not sufficiently focused. 

\begin{figure}[t]
    \centering
    \includegraphics[width=\columnwidth]{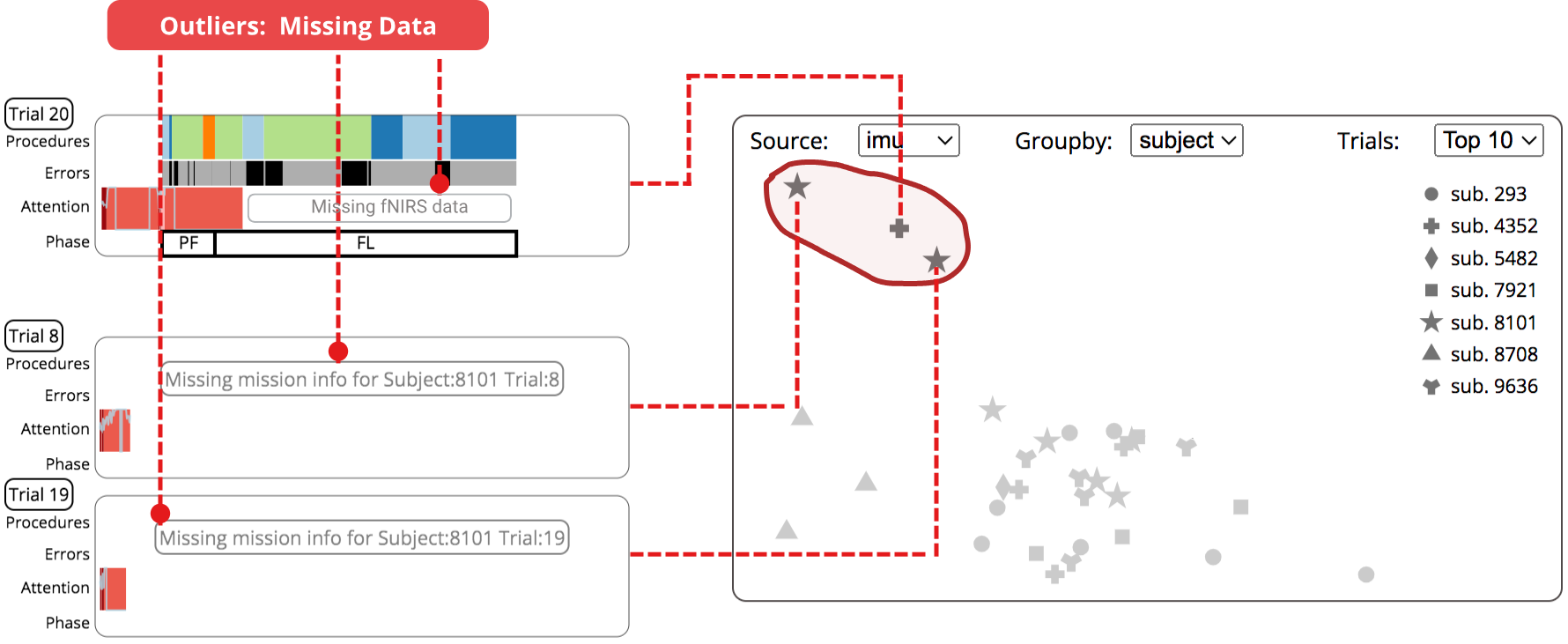}
    \vspace*{-15pt}
    \caption{Uncovering Data Quality Issues. On the right side, the scatterplot showcases sessions clustered by their IMU data, with glyphs encoding the subject ID. Variations in session counts per subject are evident, with some outliers identified in the upper left corner and highlighted through lasso selection. On the left side, the event timeline view reveals missing data points in trials 8, 19, and 20, likely attributed to mission log failures. Despite an initially comprehensive appearance, Trial 20 exhibited notable gaps in fNIRS data.}
    \label{fig:cs1_scatterplot_missing_data}
    \vspace{-.35cm}
\end{figure}

\myparagraph{Uncovering Data Quality in Flight Sessions}
In any study focused on unraveling cognitive processes, data quality plays a critical role. Acknowledging this, the trainer \revision{began the} analysis by utilizing the \revision{S}catterplot \revision{V}iew to visualize the collected data across multiple sessions. 
For this particular task, she organized the data by trial, seeking to identify \revision{sessions} where the underloaded mental state predominated. \revision{Analysis of the} scatterplot \revision{first enabled her to} identify outliers and anomalies that could indicate data quality issues, such as sensor failures or inaccuracies in data collection (see the right side of Figure~\ref{fig:cs1_scatterplot_missing_data}). The trainer investigated these outliers using the \revision{Event T}imeline \revision{V}iew, which provided a detailed breakdown of data \revision{acquired} throughout the sessions.  As shown \revision{on the left side of Figure}~\ref{fig:cs1_scatterplot_missing_data}, this examination revealed missing data points in \revision{T}rials 8, 19, and 20\revision{:} \revision{T}rials 8 and 19 \revision{only contained fNIRS data, and} lack\revision{ed} crucial information like procedures and errors, likely due to mission log failures. Meanwhile, though Trial 20 appeared comprehensive at first glance, it exhibited notable gaps in fNIRS data, implying potential technical glitches or inconsistencies in recording procedures. 
After identifying the session\revision{s} with potential issues, the trainer opted to analyze a different cluster of sessions for further examination.

\begin{figure}[t]
    \centering
    \includegraphics[width=\columnwidth]{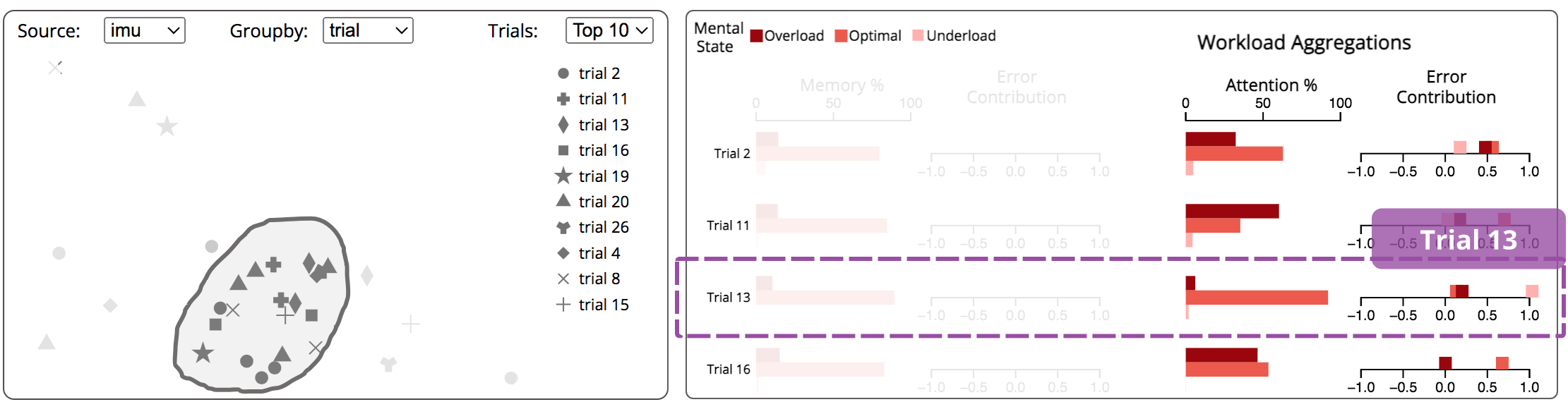}
    \vspace*{-15pt}
    \caption{Workload Aggregation View. On the left side, a selection of sessions is made within the scatterplot to identify instances where the underloaded mental state prevails. On the right side, sessions are organized by trials, depicting workload and error contribution associated with each mental state across all categories for every trial group. Notably, Trial 13 reveals a substantial correlation between errors and underload state, as indicated by the prominent pink marker near the value of 1.}
    \label{fig:cs1_error_contribution}
    \vspace{-.45cm}
\end{figure}

\myparagraph{Understanding the Link Between Errors and Underloaded Mental States}  Acknowledging that errors often signify underlying issues, \revision{the trainer} scrutinized the \revision{ Workload Aggregation View}, concentrating on sessions displaying a notable correlation between errors and underloaded mental states \revision{using the error contribution plot}. As shown in Figure \ref{fig:cs1_error_contribution}, only one session (Trial 13) out of 10 sessions exhibited significant correlations. This implies that sessions falling under this trial demonstrate a strong association between the underloaded mental state and errors. Based on these findings, the trainer \revision{selected} Trial 13 for deeper analysis. 

\begin{figure}[t]
    \centering
    \includegraphics[width=\columnwidth]{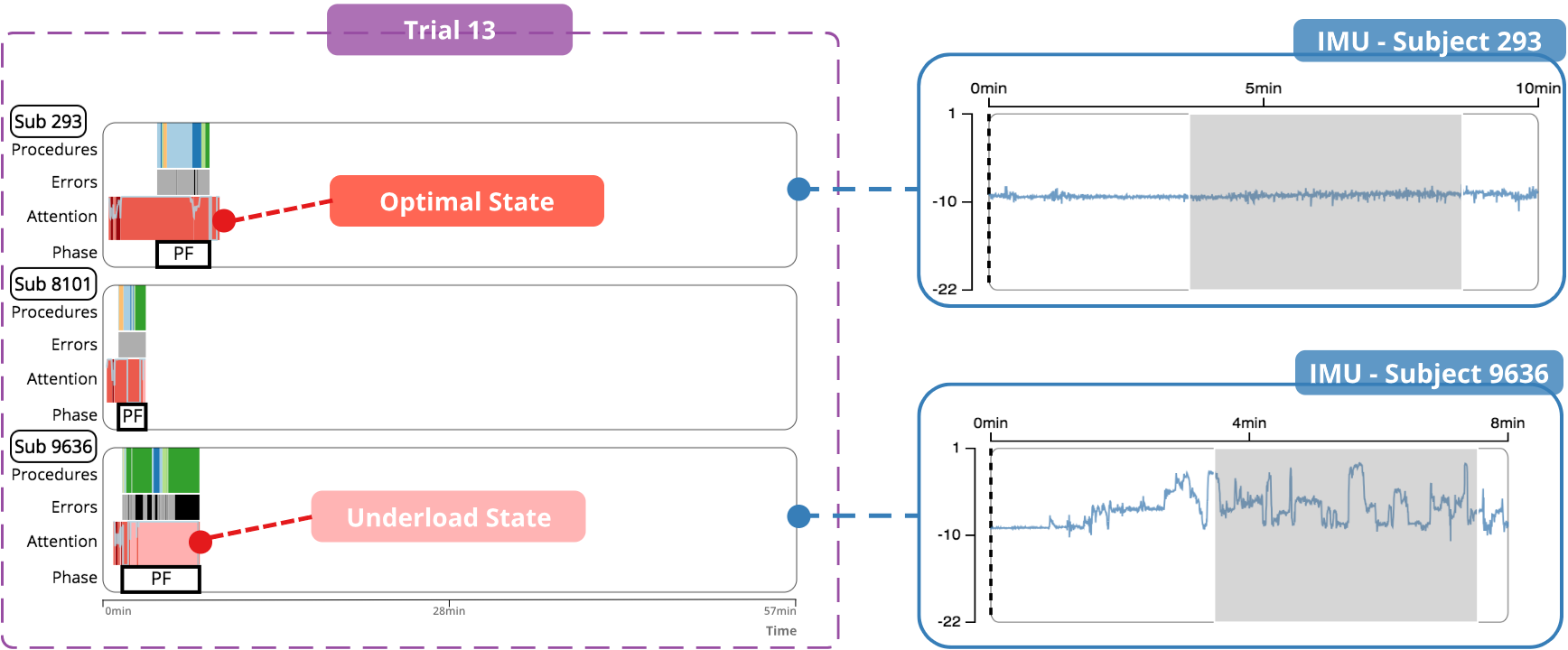}
    \caption{Event Timeline and Detail Views. On the left side, the Event Timeline View presents sessions belonging to Trial 13, conducted by different subjects. Subject 293 demonstrates sustained optimal attention and minimal errors, while Subject 9636 encounters numerous errors under an underload attention state. On the right side, the Detail View displays IMU data for Subjects 293 and 9636, revealing distinct patterns in linear acceleration. Subject 293 exhibits consistent, controlled motion, while Subject 9636 shows considerable variation, indicative of frequent stops and starts.}
    \label{fig:cs1_detail_view}
    \vspace{-.45cm}
\end{figure}

\myparagraph{Understanding \revision{Copilot} Expertise Disparities through Motion Analysis and Error Correlation}
In Trial 13, \revision{the trainer notes a} correlation between \revision{performer} errors and the underload\revision{ed} mental state, with a correlation coefficient very close to 1 (see Figure \ref{fig:cs1_error_contribution}). Upon transitioning to the \revision{Event T}imeline \revision{V}iew to analyze the sessions, the trainer quickly \revision{discovers} that Trial 13 comprises very short sessions, \revision{specifically} a task duration of under 10 minutes per subject (as shown in Figure \ref{fig:cs1_detail_view}). Examination of the phase feature in the \revision{Event T}imeline \revision{V}iew reveals that all sessions exclusively \revision{included} the preflight phase (PF), \revision{explaining} the\revision{ir} brevity.
Further scrutiny reveals that the tasks in the sessions performed by Subject 293 and Subject 9636 were completed in approximately 9 minutes. However, Subject 293 predominantly maintained an optimal attention workload state and exhibited relatively few errors, while Subject 9636 encountered numerous errors and primarily operated under an underload attention state.
To delve deeper into this discrepancy, the pilot trainer navigates the \revision{E}vent \revision{T}imeline \revision{V}iew, brushing over the entire session for Subject 293 and subsequently moves to the \revision{D}etail \revision{V}iew to assess human motion using IMU data (see the right side of Figure \ref{fig:cs1_detail_view}). Notably, Subject 293's linear acceleration plots demonstrate consistent, controlled motion, contrasting with Subject 9636's plots, which exhibit considerable variation, suggesting frequent stops and starts.
This disparity leads to the hypothesis that human motion correlates with the copilot's expertise level. To validate this conjecture, the pilot trainer reviews videos for each session, confirming her hypothesis. In the videos, it becomes evident that both subjects have a manual in front of them, but Subject 293 appears less reliant on it, whereas Subject 9636 frequently pauses to flip through the manual. This observation aligns with the notion that individuals less familiar with the task are prone to more errors and increased reliance on reference materials.

\subsection{Case Study 2: Enhancing AR Guidance Systems through User Analysis }

To showcase how \systemname facilitates the advancement of AR guidance system development, \revision{we present} two scenarios \revision{wherein} an AR guidance system developer \revision{uses} the platform. 

\begin{figure}[t]
    \centering
    \includegraphics[width=\columnwidth]{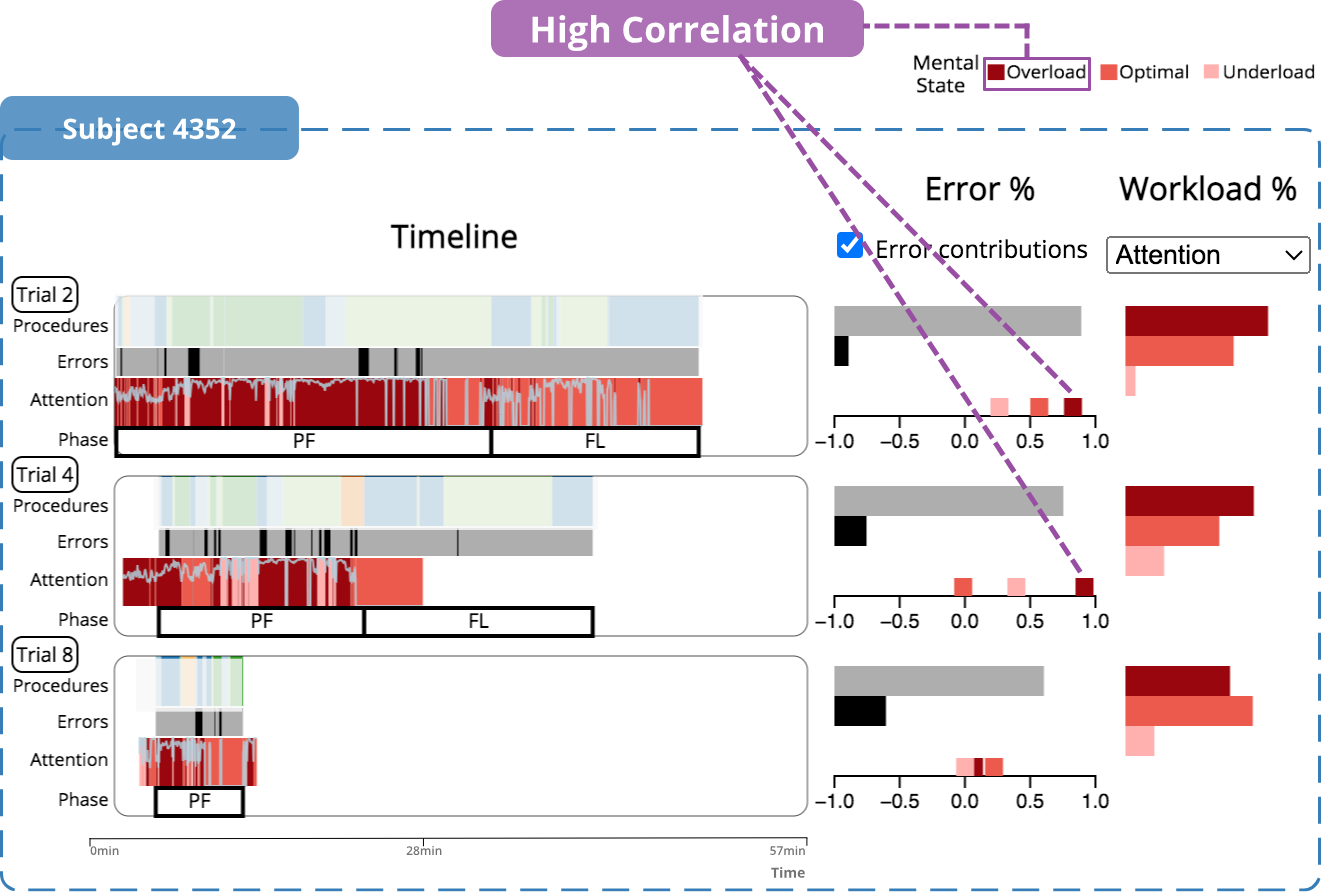}
    \vspace*{-15pt}
    \caption{The Event Timeline \revision{View} displays sessions conducted by Subject 4352 alongside error and workload summaries. The workload summary reveals a consistent overload mental state across all sessions, notably correlating with errors, particularly in Trials 2 and 4, regardless of task variations.}
    \label{fig:cs2_error_contribution}
    \vspace{-.45cm}
\end{figure}

\myparagraph{Leveraging User Profiles to Optimizing AR Flight Guidance} Understanding end-users' characteristics is paramount \revision{for effective guidance system development}. This example delves into the correlation between mental states and user \revision{characteristic} profiles, emphasizing the importance of tailoring guidance measures to assist specific user groups. To achieve this, the AR guidance system developer aims to identify emerging patterns based on pilots' performance across various tasks. Unlike \revision{the} previous case stud\revision{y}, this one \revision{focuses on the overloaded mental state, rather than the underloaded one}. \revision{T}he developer \revision{first} groups \revision{the} data by subject, assuming the issue stems from user profiles rather than the tasks themselves triggering mental states. \revision{The developer identifies S}ubjects 4352 and 293 as having significant error contributions to \revision{the} overload\revision{ed} mental state, despite both having previous piloting experience. Examining the \revision{Event T}imeline \revision{V}iew, \revision{the developer} note\revision{s} that Subject 293  and Subject 4352 completed five and three flights, respectively. Further investigation reveals that while Subject 293 displays higher overall percentages of \revision{the} overload mental state \revision{during task performance}, this condition \revision{is predominant} in only one out of \revision{their} five sessions, indicating variability in performance. Conversely, Subject 4352 consistently experiences overload across all sessions, despite task variations (see Figure \ref{fig:cs2_error_contribution}).
Furthermore, upon \revision{examining} the correlations between errors and mental states in each session conducted by Subject 4352, it becomes evident that the overload mental state exhibits a strong correlation with errors, as shown in Figure \ref{fig:cs2_error_contribution}. Examining their profiles further, Subject 293 emerges as a pilot with recent flight experience, having flown the most flights among their cohort, while Subject 4352 has been inactive in flying for 20 years. This underscores the need to consider user profiles in designing AR flight guidance systems, \revision{specifically} different system versions or adaptive features tailored to individual user profiles. 

\begin{figure*}[tb]
    \centering
    \includegraphics[width=\linewidth]{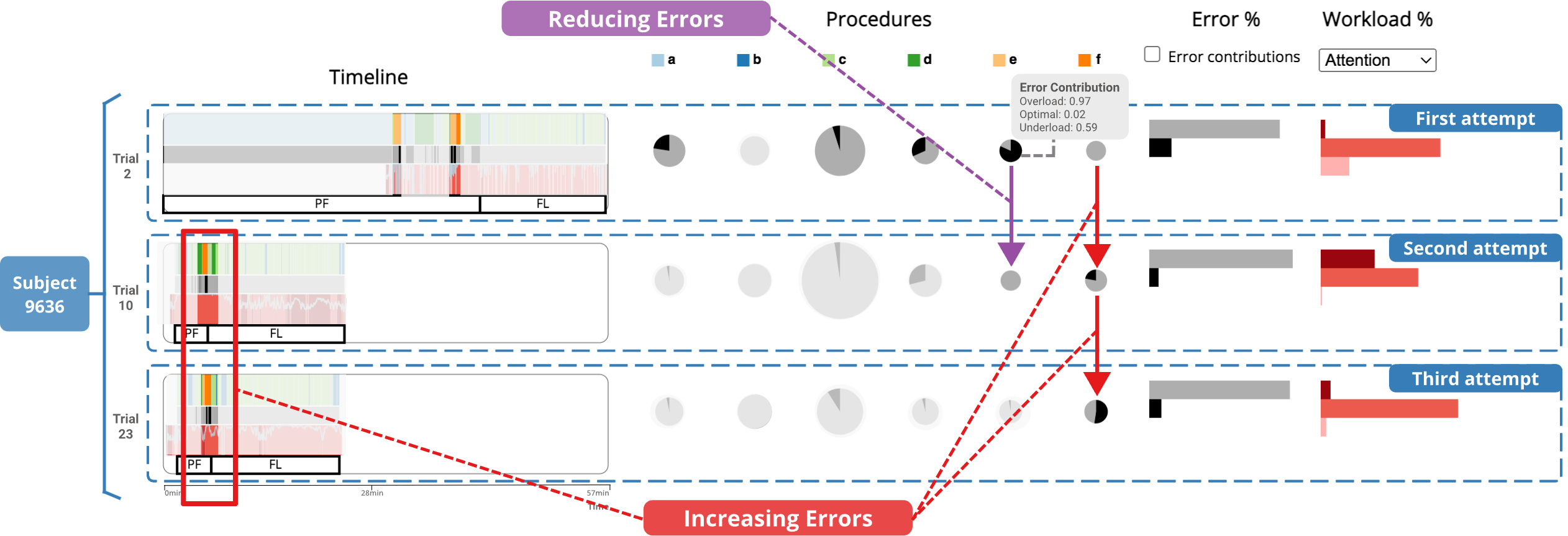}
    \caption{\revision{Performance Overview for Subject 9636. The Timeline view, Matrix view, and Workload Summaries illustrate performance across three consecutive trials (same task conditions). Notable trends include consistent task execution, reduced errors (especially in Procedure E), increased errors in Procedure F that are correlated to the transition from the preflight (PF) to flight (FL) phase, and correlations between errors and mental states, particularly in Procedure E during the first trial where the overload mental state is correlated with errors, as shown in the tooltip. Workload summaries indicate mental state improvements, with the last trial showing predominantly optimal states.}
    }
    \label{fig:cs2_enhance_novice_performance}
    \vspace{-.2cm} 
\end{figure*}

\myparagraph{Improving Performance and Mental State in AR Flight Guidance Systems}
Consider an AR guidance system developer who sets out to evaluate the progression of novice engineers over multiple flight tasks, aiming to discern the factors underlying improvement and refine guidance mechanisms to minimize errors. The \revision{developer focuses on Subject} 9636, a novice engineer, who \revision{performed} the same flight task under normal conditions \revision{three times: Trial 2, Trial 10, and Trial 23, in sequential order.}
The \revision{Event T}imeline \revision{V}iew shows Subject 9636 consistently encountered challenges during the preflight phase across all trials (as shown in Figure \ref{fig:cs2_enhance_novice_performance}). However, due to the sporadic nature of errors, pinpointing the specific procedures where the copilot struggled the most proved to be challenging. Further analysis through the \revision{Summary M}atrix \revision{V}iew revealed a consistent execution of tasks by Subject 9636 across sessions, with \revision{the most time spent on Procedure C during each session}. Notably, significant errors were observed in Procedures A, D, and E during the first \revision{attempt} (Trial 2). For Procedure E, \revision{the tooltip visualization reveals a significant correlation (0.97) between errors and the overload mental state.}
Subsequent trials displayed improvement, particularly in Procedures A and E during the second \revision{attempt} (Trial 10), where errors notably diminished, especially in Procedure E, dropping from over 70\% to zero. However, errors emerged in Procedure F during this trial. This trend persisted in the \revision{third attempt} (Trial 23), with a decline in performance in Procedure F but improvements in other procedures.
Examining the \revision{Event T}imeline \revision{V}iew provided insights into the correlation between errors in Procedure F and the transition from preflight to flight phase, suggesting the necessity for additional guidance during this phase. Furthermore, analyzing the copilot's mental state through workload summaries revealed positive impacts with improved performance. Despite high levels of underload mental state \revision{during the first attempt (Trial 2)}, subsequent trials witnessed a decrease in underload mental state, albeit accompanied by an increase in overload mental state \revision{during the second attempt (Trial 10)}. By \revision{the third attempt (Trial 23)}, the copilot achieved \revision{minimal deviations from the} optimal mental state.
These findings emphasize the interplay between overcoming flight errors and \revision{improved} copilot mental state. The developer acknowledges the imperative to focus efforts on enhancing guidance during the transition from preflight to flight to not only mitigate errors but also optimize the copilot's mental state. This case study underscores the iterative nature of analysis and adaptation essential in optimizing AR guidance systems for novice engineers' in-flight tasks.

\subsection{Expert Interview}
To validate our design decisions, we conducted a second round of interviews with \revision{five} domain experts: \revision{three} human factors and fNIRS experts (E1, E2, and E5), \revision{one} biovisualization expert (E3), and \revision{one}  neuroinformatics algorithmic design expert (E4). All of them have experience with AR-enabled applications. Four of the experts had not previously seen the tool in action (E1, E2, E4, E5), \revision{whereas} only one expert (E3) was part of the group that had previously assisted in identifying the system requirements (Section \ref{subsec:domain_requirements}). In the experiment, experts were asked to explore a group of sessions of their choice according to their interests. The fNIRS experts were specifically asked to utilize the system to gain insights into the mental states of copilots by subject. Additionally, the fNIRS experts were queried on how this tool could be integrated into their workflows to enhance efficiency. \revision{Note that fNIRS experts needed to manually synchronize different data sources, such as video and workload, to analyze this data before they had \systemname.} The design and visualization experts, on the other hand, were instructed to use the tool to explore the data with the goal of evaluating its usability.

Each interview took 50 minutes, \revision{and} began \revision{with} an overview of the project and gathering relevant background information from the participant (5 min\revision{utes}). Second, we presented our system, including a demonstration, to the participant, addressing any questions or concerns (20 min\revision{utes}). Third, participants were given the opportunity to select sessions of their preference from the Ocarina dataset (Section \ref{sec:ocarina_dataset}) to explore within the system (15 minutes). Finally, we engaged the participants in a discussion, asking questions about their initial impressions of the tool, its functionalities and features, and its potential application to their workflows (10 minutes). 

The participants were given the freedom to use \systemname and explore the available sessions. However, they were also tasked with completing three specific assignments based on the selected sessions: 1) Identify sessions demonstrating a high correlation between the overload mental state and errors, 2) Identify the most prevalent procedures within and across sessions, and 3) Utilize \systemname to interpret the sessions. They were instructed to speak while using the system, following a ``think aloud'' protocol. While the participant performed the task, an investigator took notes related to the actions performed. After completion, the participants filled out a questionnaire to express their impressions on the usability of the system. In this section, we describe the insights gathered by the participants.

\subsubsection*{Expert Insights}

\myparagraph{Data Quality Assessment} Participants also use \systemname to assess the quality of the data. In particular, E2 was very interested in ensuring that all sensors data were included before drawing any conclusions about copilot behaviors. He utilized the \revision{S}catterplot \revision{View} for this purpose. He identified outliers, hypothesizing that these sessions might have issues. Subsequently, he moved to the \revision{Event T}imeline \revision{V}iew to inspect the data confirming his hypothesis. Crucial information about procedures, errors, and flight phases was missing for the selected sessions, with only fNIRS data present. He noted that such occurrences are common due to sensor failures and expressed a desire to utilize the tool to identify such issues using the \revision{S}catterplot and \revision{Event T}imeline \revision{V}iews. Afterward, he selected another group of sessions to continue his analysis. On the other hand, E1 was not particularly focused on identifying issues in the data. While analyzing the overload mental state of Subject 4352, she observed that in trial 20, the copilot remained mostly under the optimal mental state, unlike other trials where the overload mental state was predominant. However, upon referring to the \revision{Event T}imeline \revision{V}iew, she noticed that more than half of the fNIRS data was missing for this trial. Consequently, she determined that this trial should not be included in the analysis. E3 followed a similar approach to E1.

\myparagraph{Procedures Analysis} E1, E2, and E4 \revision{extensively used the Summary Matrix View}  to \revision{identify} the predominant procedures within and across sessions for each subject. Their approach was straightforward and effective. In contrast, E3 attempted to extract this information using the \revision{Event T}imeline \revision{V}iew by comparing the duration of each procedure across the session. However, after a brief attempt, they switched to the \revision{Summary M}atrix \revision{V}iew and 
\revision{quickly}
identified the predominant procedures. E3 highlighted that the \revision{Summary M}atrix \revision{V}iew, with its normalized data across sessions, provided a clearer focus on procedures compared to the \revision{Event T}imeline \revision{V}iew. E5 took a \revision{a different approach, primarily using the \revision{Summary M}atrix \revision{V}iew to identify key procedures but also extensively employing} the \revision{Event T}imeline \revision{V}iew to observe the frequency of these procedures throughout the session, focusing on copilot performance during each procedure.

\myparagraph{Understanding Human Behavior through Error Analysis}
The majority of participants \revision{began} their analysis by examining the error contribution plot located in the \revision{Workload Aggregation View}, organizing the data by subjects rather than trials. For instance, E1 \revision{used} this view to identify subjects exhibiting a predominant overload mental state across various trials. To interpret the subjects' mental states during these trials, she navigated through the \revision{Event} Timeline \revision{V}iew and delved into the Detail \revision{V}iew. By using the \revision{D}etail \revision{V}iew, specifically the IMU signals, she observed a significant amount of human motion at the beginning of the session, transitioning to a phase characterized by consistent and controlled motion. Upon revisiting the \revision{Event} Timeline \revision{V}iew, she noted that this pattern correlated with the occurrence of errors and the flight phase, leading her to hypothesize that the pre-flight phase might be a contributing factor to errors and subsequent overload mental states due to heightened stress levels. 

\myparagraph{Managing Multimodal Data} Most participants extensively explored the various modalities available in the \systemname tool. Notably, E3 and E4 emphasized the tool's capability to visualize different data sources, including events (such as procedures and errors), fNIRS, IMU, gaze, and video, all of which were seamlessly integrated and synchronized. Among these modalities, video emerged as the preferred choice for all experts, serving as a vital resource for session analysis.
E5, in particular, heavily relied on video analysis. With a clear understanding of the conditions for each trial, E5 was keen on inspecting segments within the session where abnormal events, such as weather disturbances, occurred, to evaluate the copilot's performance. To identify these events, E5 also \revision{used} the \revision{Event T}imeline \revision{V}iew to locate procedures throughout the session. Additionally, E5 heavily relied on the IMU signal, particularly the accelerometer signal, to pinpoint segments in the video characterized by significant variance. These variations, evident through peaks and valleys in the accelerometer signal, helped identify critical moments for closer examination.

\subsubsection*{Expert Feedback}

The participants provided highly positive feedback, demonstrating interest in utilizing \systemname for their tasks and providing suggestions for enhancing the system. Following the think-aloud experiment, they were asked for any additional comments or suggestions. Here are some of their responses:
\begin{itemize}
\item E1 liked the design and interactive part of the tool. She highlighted the selection of colors to encode mental states: ``I really liked that progression (colors), a lot of people use like green, red and yellow to represent those states. And I really prefer what you guys have done, which is like the light to the dark red. I think that makes way more sense.''. She also liked the usage of pie charts: ``I really liked the use of pie charts here. I am not usually a big fan of them, but I think that that's an appropriate place for them. So I was happy to see a good pie chart.''. Regarding interactivity, she appreciated the synchronized behavior exhibited by all components:  ``I think it was both intuitive and user friendly. Being able to lasso on the scatter plots makes things really, really, really easy to capture like little clusters that you're more interested in. I liked the brushing. It was responsive on both sides of the screen (components), so I don't have to go back and forth between different sections (components) in order to look at something else, or just to switch things around.''

\item E1 also found the \revision{Event T}imeline \revision{V}iew and \revision{D}etail \revision{V}iew useful to compare and validate hypothes\revision{e}s: ``The bottom section, where I could see everything in comparison, side by side, the \revision{IMU} data with the overload\revision{/}underload state and having the video there so that you're able to validate what it is you're seeing and why, you're seeing it. I thought that was very useful.''

\item E2 liked the usage of scatterplot to detect outliers: ``The outlier detection, or the outlier capability in the upper left was kind of really powerful. I would maybe like to see that expanded from just \revision{IMU} (gaze or fNIRS) data, and maybe look at other kind of outliers, or be able to group by other kinds of data up there. So that was really useful.''. He also found the \revision{Event T}imeline \revision{V}iew very powerful: ``being able to see the procedure with the error and the workload state on the timeline view in the lower left. That was, also, I think, very helpful, really powerful, to be able to see those 3 things stacked up against each other.''

\item E3 found the system's capability to identify correlations to be effective and useful: ``I found myself working a lot with the time (timeline view), with the event sequence. Even though I know that you cannot directly compare. You know the procedures with each other because they have multiple options to do these things. It still showed me like, very well what's correlated? In which procedure didn't the error occur? And then, how was that correlated to the mental state. I think, the timeline helped a lot.'' 
E3  also liked the usage of different modalities to interpret the data: ``It was definitely cool to look into the video because you kind of wanna know what's going on. The other things are kinda abstract, and that just helps to relate a little bit to the situation. It is good to really connect what was exactly happening.''

\item E4 emphasized the \revision{Event T}imeline \revision{V}iew's capability to enable detailed segment inspection through brushing, facilitating deeper analysis: ``I like the most, was the ability to take like a section of a trial, and then like overlay that with the raw measurements of behavior such as the IMU and other markers, and the videos really nice to also see, like a raw behavior there a little bit, I really like that.''. He also appreciated the system's full interactivity: ``… each panel seems to complement each other, which is nice. I like that. I like that all the panels are tied to one another, so you can select trials in one, and then it shows it updates all the other panels and shows you nice statistics. It seems like well thought out and smooth interface.'' 

\item E5 wants to integrate \systemname in his workflow: ``part of what I do is to go through these kind of videos. Having more of that data (\systemname's features), would kind of allow me to jump to things easier … As soon as you did that (brush segments of the timeline view and synchronize this with the video), I was like, that's I wish I had that earlier.''. He also found \systemname helpful to compare different sessions: ``when you're trying to make sense of the data in your analysis, you know what you might find. For example, you know there is something significantly different between \revision{two} people or something. This tool would allow you to kind of quickly drill down into what's actually going on. Either cognitively or behaviorally. So yeah, it's helpful.''

\item \revision{Suggestions: E1 and E5 suggested enriching the Summary Matrix View, for example, by including the proportion of mental states within the pie chart associated with errors. E2 suggested support for real-time monitoring. E3 suggested the use of pattern detection to presort the sessions. Finally, E4 suggested displaying raw fNIRS data, such as an activation map for brain signals, along with the locations of fNIRS sensors.}

\vspace{-.15cm}
\end{itemize}

\subsection{Usability}

We assessed the usability of \systemname using the System Usability Score (SUS)~\cite{brooke_sus_1996}, a robust tool widely recognized for evaluating system interfaces \cite{bangor_empirical_2008}. A mean SUS score above 80 is in the fourth quartile and is acceptable. To compute the SUS, we administered a survey at the conclusion of the second interview, prompting participants to complete the standard SUS questionnaire, grading each of the 10 statements on a scale from 1 (strongly disagree) to 5 (strongly agree). The SUS grades systems on a scale between 1 and 100, and our system obtained an average score of $87 \pm 9.58$.

\section{\revision{Discussion and} Conclusion}
\label{sec:conclusion}

We presented \systemname, a novel visual analytics tool tailored for summarizing and comparing task performance sessions in Augmented Reality (AR). By integrating \revision{time series} data from fNIRS measurements, gaze, and IMU data with session logs \revision{and videos}, \systemname \revision{enables} users \revision{to} explore performer behavior and cognitive workload at various levels of granularity. \revision{Through interactive visualizations \systemname reveals patterns and anomalies in task performance, such as human errors and workload fluctuations, and their correlations with task phases. \revision{These insights} support post-hoc analysis\revision{, aiding developers in} refin\revision{ing} task guidance strategies and enhanc\revision{ing} AR-based training environments.} 

\revision{We believe \systemname integrates seamlessly into the ecosystem of AR-enabled task guidance development by enabling developers to assess the impact of different design decisions on  performer cognitive workload. For example, specific 3D interfaces designed to guide users through tasks can trigger variation in performer cognitive load depending on design. ARTiST \cite{wu_artist_2024}, for instance, leverages this by proposing a text simplification approach to reduce performer cognitive load. In turn, \systemname could facilitate more detailed exploration of the actual impact of such systems on performers. This capability could help developers create more adaptable task guidance systems that customize instructions to the performer's mental state.}

\myparagraph{Limitations} \revision{While \systemname enables users to understand the frequency and magnitude of performer movement through IMU and gaze data, it does not include an explicit representation summarizing spatial relationships between the performer and their surrounding environment. In other words, \systemname makes it easy to tell when the performer moves, but their exact pose and, in turn, action may not always be clear. This limitation persists even when IMU and gaze time series are analyzed in conjunction with egocentric video, as many AR headsets have a limited field of view which could leave important hand movements and interactions with the environment off-camera. Furthermore, \systemname does not include a visual representation of the raw data output by fNIRS sensors, instead opting for aggregated workload classification labels at each time step. While this approach enhances data interpretability for a broader audience, it may occlude anomalies in sensor performance or details that could be of interest to a brain data expert. This also creates blind reliance on the workload classifiers, with limited ability to identify potential classification errors.}

\myparagraph{Future work} \revision{First, we plan to conduct a larger user study with participants from different backgrounds to understand how well our design can adapt to new users. To intervene promptly in response to emerging issues or fluctuations in cognitive workload, we plan to 
enable real-time monitoring of task performance sessions.}
\revision{To aid in better data quality assessment and model interpretation, we also plan to explore scalable visual metaphors for analyzing fNIRS raw time series data, which may be composed of up to several dozen streams. This raw data will enable users to understand how different brain parts respond to specific stimuli and note data quality issues.}
\revision{On the machine learning front, we would like to explore techniques to automate the detection of relevant patterns and anomalies within task performance data~\cite{kloska_expert_2023}. This may include developing algorithms to classify human errors and identify optimal task guidance strategies based on historical data.} \revision{Finally, we have primarily explored the aviation domain in our use cases due to the availability of relevant data, but it is important to note that our tool is applicable across various domains, which we plan to explore in future work, as our methods work with any multimodal time series data.}
%
%

\acknowledgments{%
This work was supported by the DARPA PTG program.  Any opinions, findings, and conclusions or recommendations expressed in this material are those of the authors and do not necessarily reflect the views of DARPA.
}

\bibliographystyle{abbrv-doi}

\bibliography{bibliography}

\end{document}